\documentclass[aps,showpacs,prep,nofootinbib]{revtex4}
\usepackage{amssymb}
\usepackage{amsmath}
\usepackage{bm}

\begin{document}

\title{Gravitational waves during inflation from a 5D large-scale repulsive gravity model}
\author{$^{1}$ Luz M. Reyes , $^{1}$ Claudia Moreno \thanks{E-mail address: claudia.moreno@cucei.udg.mx},
$^{1}$ Jos\'e Edgar Madriz Aguilar \thanks{E-mail address: madriz@mdp.edu.ar} and $^{2,3}$ Mauricio Bellini
\thanks{E-mail address: mbellini@mdp.edu.ar}, }
\affiliation{$^{1}$ Departamento de Matem\'aticas, Centro Universitario de Ciencias Exactas e ingenier\'{i}as (CUCEI),
Universidad de Guadalajara (UdG), Av. Revoluci\'on 1500 S.R. 44430, Guadalajara, Jalisco, M\'exico.  \\
E-mail: edgar.madriz@red.cucei.udg.mx\\
and \\
$^{2}$ Departamento de F\'isica, Facultad de Ciencias Exactas y Naturales, Universidad Nacional de Mar del Plata (UNMdP),
Funes 3350, C.P. 7600, Mar del Plata, Argentina \\
E-mail: mbellini@mdp.edu.ar\\
$^{3}$ Instituto de Investigaciones F\'{i}sicas de Mar del Plata
(IFIMAR)- Consejo Nacional de Investigaciones Cient\'{i}ficas y
T\'ecnicas (CONICET) Argentina.}

\begin{abstract}
We investigate, in the transverse traceless (TT) gauge, the
generation of the relic background of gravitational waves,
generated during the early inflationary stage, on the framework of
a large-scale repulsive gravity model. We calculate the spectrum
of the tensor metric fluctuations of an effective 4D
Schwarzschild-de-Sitter metric on cosmological scales. This metric
is obtained after implementing a planar coordinate transformation
on a 5D Ricci-flat metric solution, in the context of a
non-compact Kaluza-Klein theory of gravity. We found that the
spectrum is nearly scale invariant under certain conditions. One
interesting aspect of this model is that it is possible to derive
the dynamical field equations for the tensor metric fluctuations,
valid not just at cosmological scales, but also at astrophysical
scales, from the same theoretical model. The astrophysical and
cosmological scales are determined by the gravity- antigravity
radius, which is a natural length scale of the model, that
indicates when gravity becomes repulsive in nature.

\end{abstract}

\maketitle

\vskip .5cm


\section{Introduction}

One of the greatest challenges of the modern cosmology is to
explain the origin of the present day accelerated expansion. In
spite of the many proposals, the problem is now still far to be
closed. Cosmologists have followed basically two lines of
research: the existence of a new exotic component in the content
of the universe and gravitation theories alternative to general
relativity. In this line of reasoning we have the well-known
$\Lambda$-CDM model in which the cosmological constant is
responsible for the acceleration in the expansion. Unfortunately,
this model suffers from the cosmological constant problem and
linked to it the cosmic coincidence problem is also a difficulty.
Quintessence models have been proposed to explain or alleviate the
cosmic coincidence problem: {\it why the value of the dark energy
density today is comparable with the present matter energy
density} \cite{intro1, intro2, intro3}. The cosmic coincidence
problem itself, motivated the appearance of interacting
quintessence models \cite{intro4, intro5, intro6}. The k-essence
models are also an alternative \cite{intro7, intro8, intro9}. None
of them are free of difficulties and thus the accelerated
expansion of the universe continue being an open problem for
theoretical cosmology. In the second line of reasoning we can find
IR modifications to general relativity, theories of gravity
in more than four dimensions, and general modified theories
of gravity \cite{intro4}. However, some of them suffer from different kinds of instabilities \cite{intro4}.\\

The extra dimensional theories have been also a recourse scenario
to address the present day accelerated expansion problem. Among
others, we can mention for instance the brane-world theories
\cite{intro10, intro11, intro12, intro13}, the M-theory
\cite{introa1,introa2}, and non-compact Kaluza-Klein (KK) theories
\cite{intro14, intro15, intro16, intro17}. The induced matter
theory is regarded a non-compact KK theory in 5D, since the fifth
extra dimension is assumed extended. This theory is mathematically
supported by the embedding Campbell-Magaard theorem \cite{cam1,
cam2}. The main idea in this theory is that matter in 4D can be
geometrically induced from a 5D Ricci-flat metric. Thus, the
theory considers a 5D geometrical vacuum defined by
$^{(5)}R_{ab}=0$, $a,b=0,1,2,3,4.$, which are the field equations
of the theory. This idea has generated a new kind of cosmological
models on which accelerating expansion periods are included. On
this setting,  has been recently introduced a new cosmological
model on which gravity manifests itself as attractive on
astrophysical scales and as repulsive on cosmological scales
\cite{ga1}. In this model our the 4D universe is represented by an
effective 4D Schwarzschild-de Sitter (SdS) metric. This metric is
obtained after the implementation of a planar coordinate
transformation on a 5D Ricci-flat SdS static metric, which is a
1-body solution to the Ricci-flat field equations of the theory.
The scalar field fluctuations of the inflaton field in an early
inflationary universe have been studied in the same physical
framework in\cite{intro18, intro19}. In these works, it is
obtained that the spectrum of the fluctuations at zeroth order is
independent of the scalar field mass on Schwarzschild scales,
while that on cosmological scales it exhibits a mass dependence
\cite{intro18, intro19}. Our theory is basically an extension of
the $\Lambda-CDM$ model and predicts an equation of state with
$\omega=-1$. However it has the virtue that can describe the
spectrum on both, cosmological and astrophysical scales. These
results indicate that in this new model the fluctuations of the
inflaton field during inflation are well described by repulsive
gravity on cosmological scales. However, as it is well known, the
inflationary mechanics in standard cosmology, also generates a
background of gravitational waves \cite{GW1}. Dark energy
cosmological scenarios have been intensivelly studied in the last
years\cite{BCNO}. The scenarios there described can explain the
generation of gravitational waves on cosmological, but not on
astrophysical scales. It is therefore useful and very instructive
to study the relic background of gravitational waves generated
during inflation on the framework of a large scale repulsive
gravity model, since gravitons under such conditions must live
imprints on the power spectrum and polarization of the cosmic
microwave background radiation (CMBR). Thus, we may have possible
ways to test this kind of large scale repulsive gravitational models.\\

In this letter we study the generation of relic background of
cosmological gravitational waves during inflation, derived from a
non-compact KK theory of gravity in the context of a repulsive
gravity theory. We have organized the letter as follows. The Sect.
II is devoted to obtain the 5D perturbed field equations. In Sect.
III we study the propagation of the 5D tensor metric fluctuations
along the extra coordinate, on cosmological scales. In Sect. IV,
we study the 4D dynamics of gravitational waves on cosmological
scales, induced from the 5D SdS spacetime and we calculate the 4D
spectrum of gravitational waves. Finally in Sect. V, we give some
final comments.

\section{The 5D perturbed field equations}

Let us begin considering in the coordinate chart $\lbrace T,R,\theta,\phi,\psi\rbrace$, the 5D Ricci-flat metric \cite{ga1}
\begin{equation}\label{a1}
dS_{5}^{2}=\left(\frac{\psi}{\psi_0}\right)^{2}\left[c^{2}f(R)dT^2-\frac{dR^2}{f(R)}-R^2(d\theta^2+sin^2\theta
d\phi^2)\right]-d\psi^2,
\end{equation}
where $f(R)=1-(2G\zeta\psi_0/Rc^2)-(R/\psi_0^2)$ is a
dimensionless metric function, $\psi$ is the spacelike and
non-compact fifth extra coordinate\footnote{In our notation
conventions henceforth, latin indices $a,b=$ run from 0 to 4,
whereas the rest of latin indices $i,j,n,l,...=$ run from 1 to
3.}. This metric is an extension to 5D spaces of the 4D SdS
metric. $T$ is a time-like coordinate, $c$ is denoting the speed
of light, $R,\theta,\phi$ are the usual spherical polar
coordinates, $\psi_0$ is an arbitrary constant with length units
and the constant parameter $\zeta$ has units of
$(mass)(length)^{-1}$. The metric (\ref{a1}) is static, however,
it can be written on a dynamical coordinate chart $\lbrace
t,r,\theta,\phi,\psi\rbrace$ by implementing the planar coordinate
transformation \cite{pc}
\begin{equation}\label{a2}
R=ar\left[1+\frac{G\zeta\psi_0}{2ar}\right]^2,\quad T=t+H\int^r dR\frac{R}{f(R)}\left(1-\frac{2G\zeta\psi_0}{R}\right)^{-1/2},
\end{equation}
$a(t)=e^{Ht}$ being the scale factor and $H$ the constant Hubble
parameter. After doing so, the line element (\ref{a1}) may be
expressed in terms of the conformal time $\tau$ in the form
\begin{equation}\label{a3}
  dS_{5}^{2} = \left(\frac{\psi}{\psi_0}\right)^2 \left[F(\tau,r)d\tau^2 - J(\tau,r)(dr^2+r^2(d\theta^2+sin^2\theta
  d\phi^2))\right]-d\psi^2,
\end{equation}
where the metric functions $F(\tau,r)$ and $J(\tau,r)$ are given by
\begin{equation}\label{a4}
F(\tau,r)=a^2(\tau)\left[1-\frac{G\zeta\psi_0}{2a(\tau)r}\right]^2\left[1+\frac{G\zeta\psi_0}{2a(\tau)r}\right]^{-2},\quad
J(\tau,r)=a^2(\tau)\left[1+\frac{G\zeta\psi_0}{2a(\tau)r}\right]^4,
\end{equation}
with $d\tau=a^{-1}(\tau)dt$ and $a(\tau)=-1/(H\tau)$, so that the constant Hubble parameter satisfies
\begin{equation}\label{a5}
H=a^{-2}\frac{da}{d\tau} .
\end{equation}
Notice that the metric in (\ref{a3}) is no more Ricci-flat. As it
was shown in \cite{ga1}, for certain values of $\zeta$ and
$\psi_0$ the metric in (\ref{a1}) has two natural horizons. The
inner one is the analogous to the Schwarzschild horizon and the
external one is the analogous to the Hubble Horizon. In the metric
in (\ref{a3}) these horizons may of course be written in the new
dynamical coordinate chart.

In order to study tensor metric fluctuations of the metric in (\ref{a3}), in the TT gauge, we use the 5D perturbed line element
\begin{equation}\label{a6}
\left.dS_5^2\right|_{pert}=\left(\frac{\psi}{\psi_0}\right)^2\left[F(\tau,x,y,z)d\tau^2-J(\tau,x,y,z)(\delta_{ij}+\Pi_{ij})dx^idx^j\right]-d\psi^2,
\end{equation}
where we have written for simplicity the 3D spatially flat line
element in cartesian coordinates,
$\Pi_{ij}(\tau,x,y,z,\psi)$ is the transverse traceless tensor that describes the tensor
metric fluctuations and satisfies: $tr(\Pi_{ij})=\Pi^{i}\,_{i}=0$ and $\partial _{i}\Pi^{ij}=0$.
Thus, we can write the spatial components of the metric as $g_{ij}=-(\psi/\psi_0)^{2}J(\tau,x,y,z,\psi)(\delta_{ij}+\Pi_{ij})$,
so that we can consider the linear approximation for the contravariant metric components $g^{ij}\simeq -(\psi/\psi_0)^{-2}J^{-1}(\tau,x,y,z,\psi)(\delta^{ij}-\Pi^{ij})$.
Of course this perturbed metric is no Ricci-flat.\\

The dynamics for the tensor modes is obtained by using the linearized 5D Einstein equations in vacuum: $\delta R_{ab}=0$, that for the case of the perturbed metric (\ref{a6}), become
\begin{eqnarray}
&&\frac{1}{4}\left(\frac{5J_{,i}F_{,j}}{J^2}-\frac{F_{,i}F_{,j}}{F^2}\right)\Pi^{ij}=0
 \label{a7}\\
&&\frac{1}{2}\left(\frac{J}{F}\right)\ddot{\Pi}_{ij}-\frac{1}{4}\left(\frac{J\dot{F}}{F^2}-\frac{3\dot{J}}{F}\right)\dot{\Pi}_{ij}-\frac{J_{,n}}{J}\delta^{ln}
\Pi_{ij,l}+\delta^{ln}\left(\frac{1}{4}\frac{F_{,l}}{F}+\frac{3}{2}\frac{J_{,l}}{J}\right)\left(\Pi_{ni,j}+\Pi_{nj,i}-\Pi_{ij,n}\right)-
\frac{1}{4}\delta^{lm}\left(\frac{J_{,l}}{J}\delta^{n} _{i}+\right. \nonumber \\
&&\left. +\frac{J_{,i}}{J}\delta^{n}_{l}-\frac{J_{,s}}{J}\delta^{ns}\delta_{il}\right)\left(\Pi_{mn,j}+\Pi_{mj,n}-\Pi_{nj,m}\right)-\frac{1}{4}
\delta^{lm}\left(\frac{J_{,j}}{J}\delta^{n}_{,l}+\frac{J_{l}}{J}\delta^{n}_{j}-\frac{J_{,s}}{J}\delta^{ns}\delta_{lj}\right)\left(\Pi_{mi,n}+\Pi_{mn,i}
-\Pi_{in,m}\right)+\nonumber \\
&& + \frac{1}{2}\delta^{ln}\left(\Pi_{ni,jl}+ \Pi_{nj,il}-\Pi_{ij,nl}\right)+ \left[\frac{1}{2}\frac{\ddot{J}}{F}-\frac{1}{4}\frac{\dot{J}\dot{F}}{F^2}
-\frac{J}{\psi_0^2}-\frac{5}{4}\frac{\dot{J}^2}{JF}-\left(\frac{J_{,nl}}{J}-\frac{J_{,n}J_{,l}}{J^2}\right)\delta^{ln}\right]\Pi_{ij}
+\left(\frac{J_{,nl}}{J}-\frac{J_{,n}J_{,l}}{J^2}\right)\delta_{ij}\Pi^{nl}-\nonumber \\
&& -\frac{1}{2}\left(\frac{F_{,l}J_{,n}}{FJ}+\frac{3J_{,l}J_{,n}}{J^2}\right)\left(\delta^{ln}\Pi_{ij}-\delta_{ij}\Pi^{ln}\right)
-\frac{1}{2}\frac{J_{,m}}{J}\left(\frac{J_{,j}}{J}\delta^{n}_{i}+\frac{J_{,i}}{J}\delta^{n}_{j}-\frac{J_{,s}}{J}\delta^{ns}
\delta_{ij}\right)\left(\delta^{lm}\Pi_{ln}-\delta_{ln}\Pi^{lm}\right)+\nonumber\\
&& +\frac{1}{2}\frac{J_{,m}}{J}\left(\frac{J_{,l}}{J}\delta^{n}_{i}+\frac{J_{,i}}{J}\delta^{n}_{l}-\frac{J_{,s}}{J}\delta^{ns}
\delta_{il}\right)\left(\delta^{lm}\Pi_{nj}-\delta_{nj}\Pi^{lm}\right) + \frac{1}{2}\frac{J_{,m}}{J}\left(\frac{J_{,j}}{J}
\delta^{n}_{l}+\frac{J_{,l}}{J}\delta^{n}_{j}-\frac{J_{,s}}{J}\delta^{ns}\delta_{lj}\right)\left(\delta^{lm}\Pi_{in}-\delta_{in}\Pi^{lm}\right)-\nonumber \\
&&-\frac{4\psi}{\psi_{0}^2}J\Pi_{ij,\psi}-\frac{1}{2}\left(\frac{\psi}{\psi_0}\right)^2J\Pi_{ij,\psi\psi}= 0,
\label{a8}
\end{eqnarray}
where the dot is denoting derivative with respect to the conformal time $\tau$. The main difference with respect
to the tensor metric fluctuation analysis usually implemented in general relativity, relies in the fact that the
equations (\ref{a7}) and (\ref{a8}) describe the dynamics of the tensor perturbations $\Pi_{ij}(\tau,x,y,z,\psi)$
on both astrophysical and cosmological scales. This feature is proper of the 5D model we are working on. By the time,
we will focus only on the dynamics at cosmological scales.

\section{Propagation of the 5D tensor fluctuations along the extra coordinate on cosmological scales}

On cosmological scales the next condition is satisfied
\begin{equation}\label{a9}
\frac{G\zeta\psi_0}{2a(\tau)r_{H}}\ll 1,
\end{equation}
where $r_{H}$ denotes the value of the radial coordinate at the horizon entry.
The perturbed field equations (\ref{a8}) with the condition (\ref{a7}) reduce in this limit case to
\begin{equation}\label{a10}
\frac{1}{2}\left(\frac{J}{F}\right)\ddot{\Pi}_{ij}-\frac{1}{4}\left(\frac{J\dot{F}}{F^2}-\frac{3\dot{J}}{F}
\right)\dot{\Pi}_{ij}-\frac{1}{2}\delta^{nl}\Pi_{ij,nl}+\left(\frac{1}{2}\frac{\ddot{J}}{F}-
\frac{1}{4}\frac{\dot{J}\dot{F}}{F^2}-\frac{J}{\psi_{0}^2}-\frac{5}{4}\frac{\dot{J}^2}{JF}\right)\Pi_{ij}
-4\frac{\psi }{\psi_0^2}J\, \Pi_{ij,\psi}-\frac{1}{2}\left(\frac{\psi}{\psi_0^2}\right)J\,\Pi_{ij,\psi\psi}=0.
\end{equation}
Notice that the left hand side of the equation (\ref{a7}) becomes identically zero on cosmological scales.
Using the fact that on this length scales $J(\tau,r)\simeq a(\tau)^2$ and $F(\tau,r)\simeq a(\tau)^2$,
the expression  (\ref{a10}) in spherical polar coordinates $(r,\theta,\varphi)$ reads
\begin{equation}\label{a11}
\ddot{\Pi}_{ij}+2{\cal H}\dot{\Pi}_{ij}-\nabla^{2}_{\bar{r}}\Pi_{ij} + 2\left(\dot{{\cal H}}
-4{\cal H}^2-\frac{a^2}{\psi_0^2}\right)\Pi_{ij}-a^{2}\left[\frac{8\psi}{\psi_0^2}\overset{\star}{\Pi}_{ij}
+\left(\frac{\psi}{\psi_0}\right)^2\overset{\star\star}{\Pi}_{ij}\right]=0,
\end{equation}
being ${\cal H}(\tau)=\dot{a}(\tau)/a(\tau)$  the conformal Hubble
parameter, $\nabla_{\bar{r}}^2$ denoting the Laplacian operator in
spherical polar coordinates and where the star $(\star)$ denotes
derivative with respect to $\psi$. Now, in order to implement the
canonical quantization of (\ref{a11}), let us introduce the
Fourier expansion
\begin{equation}\label{ac1}
\Pi^{i}_{j}(\tau,\bar{r},\psi)=\frac{1}{(2\pi)^{3/2}}\int
d^3k_r\int dk_{\psi}\sum_{\alpha
=+,\times}\!^{(\alpha)}e^{i}_{j}\left[a_{k_rk_\psi}^{(\alpha)}e^{i\bar{k}_{r}\cdot\bar{r}}\xi_{k_rk_\psi}(\tau,\psi)
+a_{k_rk_\psi}^{(\alpha)\dagger}e^{-i\bar{k}_r\cdot\bar{r}}\xi_{k_rk_\psi}^{*}(\tau,\psi)\right],
\end{equation}
with the asterisk $(*)$ denoting complex conjugate, $\alpha$ counting the number of degrees of freedom,
and where $a_{k_rk_\psi}^{(\alpha)\dagger}$ and $a_{k_rk_\psi}^{(\alpha)}$ are the creation and
annihilation operators, respectively, that satisfy the commutator algebra
\begin{eqnarray}\label{ac2}
&& \left[a_{k_rk_\psi}^{(\alpha)},a_{k_r^{\prime}k_{\psi}^{\prime}}^{(\alpha^\prime)}\right]=
g^{\alpha\alpha^{\prime}}\delta^{(3)}(\bar{k}_r-\bar{k}_{r}^{\prime})\delta(k_{\psi}-k_{\psi}^{\prime}),\\
\label{ac3} &&
\left[a_{k_rk_{\psi}}^{(\alpha)},a_{k_r^{\prime}k_{\psi}^{\prime}}^{(\alpha^{\prime})}\right]=
\left[a_{k_rk_{\psi}}^{(\alpha)\dagger},a_{k_r^{\prime}k_{\psi}^{\prime}}^{(\alpha^{\prime})\dagger}\right]=0,
\end{eqnarray}
and where $^{(\alpha)}e_{ij}$ is the polarization tensor that obeys
\begin{eqnarray}\label{ac4}
&& ^{(\alpha)}e_{ij}=\,^{(\alpha)}e_{ji},\qquad k^{i}\,^{(\alpha)}e_{ij}=0, \\
\label{ac5}
&& ^{(\alpha)}e_{ii}=0,\qquad ^{(\alpha)}e_{ij}(-\bar{k}_r)=\,^{(\alpha)}e_{ij}^{*}(\bar{k}_r).
\end{eqnarray}
In this manner, substituting (\ref{ac1}) in (\ref{a11}) we obtain
\begin{equation}\label{ac6}
\ddot{\xi}_{k_rk_{\psi}}+2{\cal H}\dot{\xi}_{k_rk_{\psi}}+\left[k_r^2+2(\dot{{\cal H}}-4{\cal H}^2
-\psi_{0}^{-2}a^2)\right]\xi_{k_rk_{\psi}}-a^2\left[\frac{8\psi}{\psi_0^2}\,\overset{\star}{\xi}_{k_{r}k_{\psi}}
+\left(\frac{\psi}{\psi_0}\right)^2\overset{\star\star}{\xi}_{k_rk_{\psi}}\right]=0,
\end{equation}
which is the equation that describes the dynamics of the 5D tensor modes $\xi_{k_rk_\psi}(\tau,\psi)$.\\

We decompose the 5D tensor modes $\xi_{k_rk_\psi}(\tau,\psi)$ into
the KK-modes
\begin{equation}\label{ac7}
\xi_{k_rk_{\psi}}(\tau,\psi)=\int d\tilde{m}\, \xi_{\tilde{m}}(\tau)\Omega_{\tilde{m}}(\psi),
\end{equation}
and the equation (\ref{ac6}) reduces to the system
\begin{eqnarray}\label{ac8}
&& \ddot{\xi}_{\tilde{m}}+2{\cal H}\dot{\xi}_{\tilde{m}}+\left[k_r^2+2(\dot{\cal H}-4{\cal H}^2-\psi_0^{-2}a^2)+\tilde{m}^2\right]\xi_{\tilde{m}}=0,\\
\label{ac9}
&& \left(\frac{\psi}{\psi_0}\right)^2\overset{\star\star}{\Omega}_{\tilde{m}}+\frac{8\psi}{\psi_0^2}\overset{\star}{\Omega}_{\tilde{m}}+\tilde{m}^2\Omega_{\tilde{m}}=0,
\end{eqnarray}
where the parameter $\tilde{m}^2$ corresponds to the square of the
KK-mass measured by a 5D observer. To simplify the structure of
(\ref{ac9}) we introduce the field transformation:
$\Omega_{\tilde{m}}(\psi)=(\psi_0/\psi)^4P_{\tilde{m}}(\psi)$, and
the equation (\ref{ac9}) results to be
\begin{equation}\label{ac10}
\psi^2\overset{\star\star}{P}_{\tilde{m}}-(12-\tilde{m}^2\psi_0^2)P_{\tilde{m}}=0.
\end{equation}
After resolve this equation, we obtain
\begin{equation}\label{a14}
P_{\tilde{m}}(\psi)=C_1\psi^{\gamma_1}+C_2\psi^{\gamma_2},
\end{equation}
where $\gamma_{1}=(1/2)[1+\sqrt{49-4\tilde{m}^2\psi_0^2}]$, $\gamma_2=(1/2)[1-\sqrt{49-4\tilde{m}^2\psi_0^2}]$ and $C_1$, $C_2$ are integration constants.\\

In order to study the stability of the KK-modes, let us rewrite
the equation (\ref{ac9}) in terms of the conformal spatial fifth
coordinate: $u=\ln(\psi/\psi_0)$. Using the auxiliary field
transformation:
$\Omega_{\tilde{m}}(u)=e^{-(7/2)u}L_{\tilde{m}}(u)$, we obtain
\begin{equation}\label{ac12}
\frac{d^2L_{\tilde{m}}}{du^2}+\left(\tilde{m}^2\psi_0^2-\frac{49}{4}\right)L_{\tilde{m}}=0.
\end{equation}
The stability of the KK-modes depends directly on the stability of
the solutions of the equations (\ref{ac8}) and (\ref{ac9}).  One
can see from (\ref{ac12}) that for $\tilde{m}^2\psi_0^2>49/4$, the
redefined modes $L_m$ are coherent (stable) on the ultraviolet
(UV) sector. When $\tilde{m}^2\psi_0^2<49/4$ those $L_m$-modes are
unstable and diverge when $u$ tends to infinity. The $L_m$- mode
with $\tilde{m}=0$ is stable. However, notice that even when the
$L_m$-modes are unstable for $\tilde{m}^2\psi_0^2<49/4$, the modes
$\Omega_{\tilde{m}}(u)=e^{-(7/2)u}L_{\tilde{m}}(u)$, for any
$\tilde{m}$ real, never diverges. On the other hand, in the case
of the equation (\ref{ac8}), using the transformation
$\xi_{\tilde{m}}=\Theta_{\tilde{m}}\exp(-\int{\cal H}d\tau)$, we
obtain
\begin{equation}\label{ultt1}
\ddot{\Theta}_{\tilde{m}}+[k_r^2+\dot{\cal H}-9{\cal H}^2-2\psi_{0}^{-2}a^2+\tilde{m}^2]\Theta_{\tilde{m}}=0.
\end{equation}
It follows from this equation that the modes $\Theta_{\tilde{m}}$ are stable for
\begin{equation}\label{ultt2}
k_r^2>9{\cal H}^2+2\psi_{0}^{-2}a^2-\dot{{\cal H}}-\tilde{m}^2\geq 0.
\end{equation}
This condition is satisfied for $\tilde{m}\ll 1$. However, for
modes with a bigger mass $\tilde{m}$ and $k_r^2\ll 1$ the
condition (\ref{ultt2}) is no valid and hence, on very large
scales, the modes $\Theta_{\tilde{m}}$ are unstable.

\section{The 4D induced dynamics of  gravitational waves on cosmological scales}

In order to derive the 4D dynamics of the inflationary universe on
cosmological scales induced by the 5D field equations (\ref{a11}),
let us assume that the 5D spacetime can be foliated by a family of
hypersurfaces $\Sigma:\psi=\psi_0$, where our usual 4D universe is
represented by a leaf member of the foliation $\Sigma$. On every
leaf the metric induced by (\ref{a3}) take the form
\begin{equation}\label{a15}
dS_{4}^{2}=F(\tau,r)dr^2-J(\tau,r)[dr^2 +r^2(d\theta^2 +sin^2\theta d\phi^2)],
\end{equation}
where the metric functions $F(\tau,r)$ and $J(\tau,r)$ are now
given in terms of the physical mass $m=\zeta\psi_0$ (introduced by
the first time in \cite{ga1}) as it is shown in the expressions
\begin{equation}\label{a16}
F(\tau,r)=a^2(\tau)\left[1-\frac{Gm}{2a(\tau)r}\right]^{2}\left[1+\frac{Gm}{2a(\tau)r}\right]^{-2},\quad J(\tau,r)=a^{2}(\tau)\left[1+\frac{Gm}{2a(\tau)r}\right]^{4},
\end{equation}
which are valid for $r>Gm/(2a)$. The induced metric (\ref{a15})
describes a black hole immersed in an expanding universe, where
the expansion is driven by a kind of cosmological constant, whose
value depends of the $\psi_0$-value, which is related to the
Hubble constant through: $\psi_0=c^2/H$, with $c$ denoting the
speed of light \cite{ga1}. The condition that determines
cosmological scales (\ref{a9}), can be now expressed in terms of
this physical mass $m$, in the form
\begin{equation}\label{ab16}
\frac{Gm}{2a(\tau)r_{H}}\ll 1,
\end{equation}
so that in this limit the functions: $\left.F(\tau,r)\right|_{r\gg
Gm/(2a)} \simeq F(\tau) = a^2(\tau)$,
$\left.J(\tau,r)\right|_{r\gg Gm/(2a)} \simeq J(\tau) =
a^2(\tau)$, become independent of $r$, and describe an isotropic
and asymptotically homogenous expanding universe with a scale
factor $a(\tau)$, in agreement with that one expects in a FRW
cosmology. In particular, when $a(\tau)=-1/(H\tau)$, we are
dealing with a de Sitter (inflationary expansion) with an equation
of state $p/\rho=\omega=-1$. This topic was studied with detail in
\cite{intro19}. The realization of a reheating epoch after
inflation from this asymptotic de Sitter metric, with particles
and gravitons creation included, was studied in\cite{AB}. \\

As it was shown in \cite{ga1}, a very interesting aspect of
(\ref{a15}), is the existence of a length scale that separates
regions where gravity is attractive (small astrophysical scales),
from regions where it behaves as repulsive (cosmological scales).
This length scale is known as the gravitational-antigravitational
radius, which in the coordinate chart $(T,R)$ is determined by the
expression
\begin{equation}\label{a17}
R_{ga}=(Gm\psi_0^2)^{1/3}.
\end{equation}
In the dynamical coordinate chart $(\tau,r)$, we have a new gravitational-antigravitational radius $r_{ga}$ given by
\begin{equation}\label{a18}
r_{ga}=\frac{1}{2a(\tau)}\left[R_{ga}-Gm+\sqrt{R_{ga}^2-2GmR_{ga}}\,\right].
\end{equation}
It can be easily seen from this expression that  $r_{ga}$ is
positive when $R_{ga}\ge 2Gm$, and due to (\ref{a17}) such a
condition can be written as
\begin{equation}\label{a19}
\epsilon=\frac{mH}{M_p^2}\leq \frac{1}{2\sqrt{2}}\simeq 0.353553,
\end{equation}
where we have used $G=M_p^{-2}$. It follows from this last condition that the value of the fifth coordinate $\psi_0$ is restricted by: $\psi_0\ge (2\sqrt{2}m)/M_p^2$. \\

In order to derive the 4D field equations induced on $\Sigma$ by
the 5D geometry, now we consider a separable 5D field
\begin{equation}\label{ada1}
 \Pi_{ij}(\tau,\bar{r},\psi)= h_{ij}(\tau,\bar{r})\Gamma(\psi).
\end{equation}
Thus, the part of (\ref{a11}) that depends only on the extra coordinate $\psi$, has the form
\begin{equation}\label{ad1}
\left(\frac{\psi}{\psi_0}\right)^2\overset{\star\star}{\Gamma}+\frac{8\psi}{\psi_0^2}\overset{\star}{\Gamma}-\lambda_{0}\Gamma=0,
\end{equation}
$\lambda_0$ being a separation constant. The general solution of
this equation reads
\begin{equation}\label{ad2}
\Gamma(\psi)=B_{1}\psi^{\alpha_1}+B_2\psi^{\alpha_2},
\end{equation}
where $B_1$ and $B_{2}$ are integration constants and,
$\alpha_1=(1/2)(-7+\sqrt{49+4\lambda_{0}\psi_{0}^2})$,
$\alpha_2=(1/2)(-7-\sqrt{49+4\lambda_0\psi_0^2})$. With the help
of (\ref{ada1}), it follows from (\ref{a11}) that the 4D induced
field equations on $\Sigma:\psi=\psi_0$ have the form
\begin{equation}\label{a20}
\ddot{h}_{ij}+2{\cal H}\dot{h}_{ij}-\nabla_{\bar{r}}^2h_{ij}+\left[2(\dot{{\cal H}}-4{\cal H}^2)-\frac{\sigma}{\psi_{0}^2}a^2\right]h_{ij}=0,
\end{equation}
where $\sigma=(2\psi_0^2+\gamma_0)\psi_0^2$ with
$\gamma_0=[(8/\psi_0)(\overset{\star}{\Omega}/\Omega)+(\overset{\star\star}{\Omega}/\Omega)]|_{\psi=\psi_0}$,
which for $B_2=0$ in (\ref{ad2}), reduces to
$\sigma=2+\alpha_{1}(\alpha_{1}-7)$. It is important to notice
that the last term on the left hand side of (\ref{a20}) is a new
correction coming from the 5D geometry.  If we introduce the
auxiliary field $w_{ij}$ defined by
$h_{ij}(\tau,r)=a^{-1}(\tau)w_{ij}(\tau,r)$, the equation
(\ref{a20}), yields
\begin{equation}\label{a21}
\ddot{w}^{i}_{j}-\nabla^2w^{i}_{j}+\left(\dot{{\cal H}}-9{\cal H}^{2}-\frac{\sigma}{\psi_0^2}a^2\right)w^{i}_{j}=0.
\end{equation}
To quantizate the field equation (\ref{a21}), we use the 5D
Fourier expansion (\ref{ac1}) specialized on $\Sigma$
\begin{equation}\label{a22}
w^{i}_{j}(\tau,\bar{r})=\frac{1}{(2\pi)^{3/2}}\int d^{3}k_{r}
dk_{\psi}\sum_{\alpha=+,\times}\,^{(\alpha)}\!e^{i}_{j}\left[a_{k_r
k_{\psi}}^{(\alpha)}e^{i\bar{k}_{r}\cdot\bar{r}}\xi_{k_r
k_{\psi}}(\tau,\psi) + a_{k_r
k_{\psi}}^{(\alpha)\dagger}e^{-i\bar{k}_r\cdot\bar{r}}\xi_{k_r
k_\psi}^{*}(\tau,\psi)\right]\delta(k_{\psi}-k_{\psi_0}),
\end{equation}
whose integration on $k_{\psi}$ yields the 4D expansion
\begin{equation}\label{a23}
w^{i}_{j}(\tau,\bar{r})=\frac{1}{(2\pi)^{3/2}}\int
d^{3}k_{r}\sum_{\alpha=+,\times}\,^{(\alpha)}\!e^{i}_{j}\left[a_{k_r}^{(\alpha)}e^{i\bar{k}_{r}\cdot\bar{r}}\xi_{k_r}(\tau)
+
a_{k_r}^{(\alpha)\dagger}e^{-i\bar{k}_{r}\cdot\bar{r}}\xi_{k_r}^{*}(\tau)\right],
\end{equation}
where we have done the formal identifications:
$\xi_{k_r}(\tau)=\xi_{k_r k_{\psi_0}}(\tau)$,
$a_{k_r}^{(\alpha)}=a_{k_r k_{\psi_0}}^{(\alpha)}$,
$a_{k_r}^{(\alpha)\dagger}=a_{k_r k_{\psi_0}}^{(\alpha)\dagger}$,
with the creation and annihilation operators
$a_{k_r}^{(\alpha)\dagger}$ and $a_{k_r}^{(\alpha)}$,
respectively. These operators satisfy the algebra
\begin{eqnarray}
&&\left[a_{k_r}^{(\alpha)},a_{k_{r}^{\prime}}^{(\alpha^{\prime})\dagger}\right]=g^{\alpha\alpha^{\prime}}\delta^{(3)}\left(\bar{k}_r-\bar{k}^{\prime}_{r}\right),\label{a24}\\
&& \left[a_{k_r}^{(\alpha)},a_{k_{r}^{\prime}}^{(\alpha^{\prime})}\right]=\left[a_{k_r}^{(\alpha)\dagger},a_{k_{r}^{\prime}}^{(\alpha^{\prime})\dagger}\right]=0,\label{a25}
\end{eqnarray}
and where the same properties for the polarization tensor $_{(\alpha)}e^{ij}$, given by (\ref{ac4}) and (\ref{ac5}), hold.\\

The quantization procedure requires that the modes $\xi_{k_r}$ must satisfy the commutation relation
\begin{equation}\label{a28}
\left[\xi_{k_r}(\tau,\bar{r}),\dot{\xi}_{k_{r}^{\prime}}(\tau,\bar{r}^{\prime})\right]=i\delta^{(3)}\left(\bar{r}-\bar{r}^{\prime}\right).
\end{equation}
Therefore, with the use of (\ref{a23}), the expression (\ref{a28})
complies with the normalization condition
\begin{equation}\label{a29}
\xi_{k_r}\dot{\xi}_{k_r}^{*}-\xi_{k_r}^{*}\dot{\xi}_{k_r}=i,
\end{equation}
on the UV-sector. In this manner, if the modes $\xi_{k_r}$ satisfy
(\ref{a29}), automatically (\ref{a28}) will hold. After inserting
(\ref{a23}) in (\ref{a21}), we obtain
\begin{equation}\label{a30}
\ddot{\xi}_{k_r}+\left[k_{r}^2-\left(9{\cal H}^2-\dot{{\cal H}}+\frac{\sigma}{\psi_0^2}a^2\right)\right]\xi_{k_r}=0.
\end{equation}
The general solution for this equation is given by
\begin{equation}\label{a31}
\xi_{k_r}(\tau)=A_{1}\sqrt{-\tau}{\cal H}_{\nu_T}^{(1)}(-k_{r}\tau)+A_{2}\sqrt{-\tau}{\cal H}_{\nu_T}^{(2)}(-k_r\tau),
\end{equation}
${\cal H}_{\nu_T}^{(1,2)}$ being the first and second kind Hankel
functions with order
$\nu_{T}=[1/(2\psi_0H)]\sqrt{33\psi_{0}^{2}H^2+4\sigma}$. After
make the Bunch-Davies normalization \cite{Bunch}, we obtain the
normalized solution that satisfies (\ref{a29}) in the UV-sector.
It is
\begin{equation}\label{a32}
\xi_{k_r}(\tau)=i\frac{\sqrt{\pi}}{2}\sqrt{-\tau}{\cal H}_{\nu_T}^{(2)}[-k_r\tau].
\end{equation}
Once we have calculated the analytic expression for the quantum
tensor modes $\xi_{k_r}(\tau)$, we are now in position to obtain
the spectrum of gravitational waves induced on our 4D spacetime
$\Sigma$.

\subsection{The 4D  spectrum of gravitational waves}

In order to obtain the 4D spectrum of gravitational waves on
Super-Hubble (cosmological) scales, let us calculate the amplitude
of the 4D tensor metric fluctuations on the IR-sector
($-k_r\tau\ll 1$), which is given by
\begin{equation}\label{a33}
\left<h^{2}(\tau)\right>_{IR}=\frac{4a^{-2}(\tau)}{\pi^2}\int_{0}^{\epsilon k_{H}}
\frac{dk_r}{k_r}k_{r}^{3}\left.\left[\xi_{k_r}(\tau)\xi_{k_r}^{*}(\tau)\right]\right|_{IR},
\end{equation}
where $\epsilon=k_{max}^{IR}/k_{p}\ll 1$ is a dimensionless
parameter, being $k_{max}^{IR}=k_{H}(\tau_i)=\sqrt{9{\cal
H}^{2}-\dot{{\cal
H}}+(\sigma/\psi_0^2)a^2}=\sqrt{10+[\sigma/(\psi_0^2H^2)]}\tau^{-1}$
the wave number related to the Hubble radius at the conformal time
$\tau_i$, which corresponds to the time when the modes re-enter to
the horizon, and $k_p$ is the Planckian wave number. During
inflation $H=0.5\cdot 10^{-9}\,M_p$ and thus the values of
$\epsilon$ range from $10^{-5}$ to $10^{-8}$, corresponding to the
number of $e$-foldings: $N_e=63$ \cite{AAraya}. Considering the
asymptotic expansion for the Hankel function ${\cal
H}_{\nu_T}^{(2)}[x]\simeq -(i/\pi)\Gamma(\nu_T)(x/2)^{-\nu_T}$,
the expressions (\ref{a32}) and (\ref{a33}) yield
\begin{equation}\label{a34}
\left<h^{2}(\tau)\right>_{IR}=\frac{2^{2\nu_T-1}}{\pi^3}H^2\Gamma^{2}(\nu_T)(-\tau)^{3-2\nu_T}\int_{0}^{\epsilon k_H}\frac{dk_r}{k_r} k_r^{3-2\nu_T}.
\end{equation}
Performing the integration in (\ref{a34}) we arrive to
\begin{equation}\label{a35}
\left<h^2(\tau)\right>_{IR}=\frac{2^{2\nu_T-1}}{\pi^3}\frac{\Gamma^2(\nu_T)}{3-2\nu_T}H^2\epsilon^{3-2\nu_T}(-k_r\tau)^{3-2\nu_T},
\end{equation}
which in terms of the cosmic time $t$ can be written in the form
\begin{equation}\label{a36}
\left<h^2(t)\right>_{IR}=\frac{2^{1+2\nu_T}}{\pi}\frac{\Gamma^2(\nu_T)}{3-2\nu_T}\left(\frac{H}{2\pi}\right)^2\epsilon^{3-2\nu_T}\left(\frac{k_H}{aH}\right)^{3-2\nu_T}.
\end{equation}
The power spectrum for gravitational waves can be then extracted from (\ref{a36}), resulting
\begin{equation}\label{a37}
{\cal P}_{g}(k_H)=\frac{2^{1+2\nu_T}}{\pi}\Gamma^2(\nu_T)\left(\frac{H}{2\pi}\right)^{2}\left.\left(\frac{k_r}{aH}\right)^{3-2\nu_T}\right|_{k_H}.
\end{equation}
This spectrum is nearly scale invariant ($\nu_T\simeq 3/2$) for
$\sigma\simeq -6\psi_0^2H^2$. This value of $\sigma$ corresponds
to $\alpha_1\simeq \frac{7}{2}\pm\sqrt{41-24\psi_0^2H^2}$. Notice
that the last term in the brackets of (\ref{a20}): $-{4\sigma\over
\psi^2_0 H^2}$, is the responsible to broke the scale invariance
in the spectrum of gravitational waves. In our formalism this term
appears due to the existence of the extra non-compact space-like
dimension. However, in the standard 4D inflationary theory, a
possible origin of this term could be in the coupling of the
inflaton to new sectors of light degrees of freedom which are
closely spaced along its trajectory\cite{SSZ}. From WMAP5\cite{D}
data, we obtain that the acceptable values of $-{\sigma\over
\psi^2_0 H^2}$ is in the range
\begin{equation}
5.89649 > -{\sigma\over \psi^2_0 H^2} > 5.7659,
\end{equation}
once we require that the spectral index to be in the range: $n_s =
0.963^{+0.014}_{-0.015}$, for a tensor index $n_T=1-n_s$. In the
same context, it can be shown that the spectral index for scalar
metric fluctuations is given by: $n_s\simeq 4-2\nu_s\simeq
4-\sqrt{9+16k_{\psi_{0}}^2H^{-2}}$. If we define, in analogy with
the definition of slow-roll parameters of 4D inflationary models,
the parameter $\theta = 16k_{\psi_0}^2/H^2\ll 1$, the relation
$2-\nu_s\simeq\theta$ holds. Hence, the scalar-tensor ratio
$r_{ST}=A_{T}^2(k_r)/A_{S}^2(k_r)$ satisfies: $r_{ST}\simeq
(1/8)\theta\simeq [k_{\psi_0}^2/(2H^2)]$. Thus, if the parameter
$k_{\psi_0}$ varies in the interval $0<k_{\psi_0}^2\leq
(0.15)^2H^2$, then the scalar-tensor ratio ranges in the interval
$0<r_{ST}\leq 0.0112$ \cite{AB,ur1,ur2}. With PLANCK experiment,
through its B-mode polarization of relic gravitational waves
observations, will be able to reject models with $r_{ST}>0.095$
\cite{BM}. However, the restrictions on  $k_{\psi_0}^2$ must be
more carefully established in order to have a more accurate
prediction for the scalar-tensor ratio $r_{ST}$ in this
large-scale repulsive-gravity model, but it goes beyond the scope
of this paper.

\section{ Final Comments}

In this letter we have studied the relic background of
gravitational waves generated during inflation, on the framework
of a cosmological model in which gravity manifests as attractive
on astrophysical
scales and as repulsive on cosmological scales. We have obtained particularly
the spectrum of gravitational waves on cosmological scales.\\

We use a static and Ricci-flat 5D metric to derive, via a planar
coordinate transformation, a dynamical 5D metric that can be
considered as a 5D extension of a Schwarzschild-de-Sitter metric.
The fifth extra dimension is considered as non-compact and our
universe is represented by a 4D spacetime locally and
isometrically embedded into a 5D ambient space metrically
described by the field equations: $^{(5)}R_{ab}=0$. The
cosmological and astrophysical scales are, in this model, defined
in terms of a length scale called the gravity-antigravity radius
$r_{ga}$ \cite{ga1}. Thus, the astrophysical scale, normally used
in astrophysical models, coincides with the region defined by
$r<r_{ga}$ and gravity is attractive here, whereas the
cosmological scale coincides with the region $r>r_{ga}$ in which
gravity is repulsive in nature \cite{ga1,intro18,intro19}.
Dynamical 5D field equations for the tensor metric fluctuations in
the TT-gauge of the non-static metric (\ref{a3}) are obtained (see
the equations (\ref{a7}) and (\ref{a8})). The interesting of these
equations is that they are valid for both astrophysical and
cosmological scales. In 4D the induced effective field equations
contain an additional term which depends of a constant parameter
$\sigma$ (see the expression (\ref{a20}) and the definitions below
it). This additional term is a new contribution coming from the 5D
geometry. The spectrum of gravitational waves, given by the
equation (\ref{a37}), resulted nearly scale invariant when the
parameter $\sigma$ in the extra term of (\ref{a20}) satisfies
$\sigma\simeq -6\psi_{0}^2H^2$. The amplitude of the 4D square
tensor metric fluctuations on the IR-sector, given by (\ref{a36}),
behaves in a similar manner as they do in standard 4D inflationary
models. We consider that these results are of great utility to
test the viability of this new class of large-scale repulsive
gravity models.

\section*{Acknowledgements}

\noindent L. M. Reyes thanks Mathematics Department of CUCEI-UdG for kind hospitality. C. Moreno and
J. E. Madriz-Aguilar  acknowledge CONACYT (Mexico) and Mathematics Department of CUCEI- UdG for financial
support. M. Bellini acknowledges CONICET (Argentina) and UNMdP for financial support.


\bigskip

\begin{thebibliography}{99}
\bibitem{intro1} L. M. Reyes and J. E. Madriz-Aguilar, Class. Quant. Grav. {\bf 28} (2011) 215007.
\bibitem{intro2} Y-F Cai, E. V. Saridakis, M. R. Setare and J. P. Xia, Phys. Rep. {\bf 493} (2010) 1-60 (arXiv: 0909.2776 [hep-th]).
\bibitem{intro3} E. V. Linder, Gen. Rel. Grav. {\bf 40} (2008) 329-356.
\bibitem{intro4} D. Sapone, Int. J. Mod. Phys. A {\bf 25} (2010) 5253-5331.
\bibitem{intro5} W. Zimdahl and D. Pavon, Phys. Lett. B {\bf 521} (2001) 133-138.
\bibitem{intro6} L. P. Chimento, A. S. Jakubi, D. Pavon and W. Zimdahl, Phys. Rev. D {\bf 67} (2003) 083513 (arXiv: astro-ph/0303145).
\bibitem{intro7} C. Armendariz-Picon, V. F. Mukhanov and P. J. Steinhardt, Phys. Rev. Lett. {\bf 85}, (2000) 4438 (arXiv: astro-ph/0004134).
\bibitem{intro8} C. Armendariz-Picon, V. F. Mukhanov and  P. J. Steinhardt, Phys. Rev. D {\bf 63} (2001) 103510.
\bibitem{intro9} R. de Putter and E. V. Linder, Astropart. Phys. {\bf 28} (2007) 263-272.
\bibitem{intro10} N. Arkani-Hamed, S. Dimopoulos and G. Dvali, Phys. Lett. B {\bf 429} (1998) 263.
\bibitem{intro11} I. Antoniadis, N. Arkani-Hammed, S. Dimopoulos and G. Dvali, Phys. Lett. B {\bf 436} (1998) 257.
\bibitem{intro12} L. Randall and R. Sundrum, Phys. Rev. Lett. {\bf 83} (1999) 3370.
\bibitem{intro13} L. Randall and R. Sundrum, Phys. Rev. Lett. {\bf 83} (1999) 4690.
\bibitem{introa1} M. J. Duff, Int. J. Mod. Phys. A {\bf 11} (1996) 5623-5642.
\bibitem{introa2} A. Miemiec and I. Schnakenburg, Forsch. Phys. {\bf 54} (2006) 5-72 (arXiv: hep-th/0509137).
\bibitem{intro14} P. S. Wesson, Space-Time-Matter (Singapore: World Scientific, 1999).
\bibitem{intro15} J. M. Overduin and P. S. Wesson, Phys. Rep. {\bf 283} (1997) 302.
\bibitem{intro16} P. S. Wesson, Five Dimensional Physics  (Singapore: World Scientific, 2006).
\bibitem{intro17} J. E. Madriz-Aguilar and M. Bellini, Phys. Lett. B {\bf 596} (2004) 116-122 (arXiv: gr-qc/0405024).
\bibitem{cam1} J. E. Campbell, A course of Differential Geometry, Claredon, Oxford, 1926.
\bibitem{cam2} E. Anderson, F. Dahia, J. Lidsey and C. Romero, J. Math. Phys. {\bf 44} (2002) 5108.
\bibitem{ga1} J. E. Madriz-Aguilar and M. Bellini, Phys. Lett. {\bf B679} (2009) 306-310.
\bibitem{intro18} L. M. Reyes, J. E. Madriz-Aguilar and M. Bellini, Eur. Phys. J. Plus {\bf 126} (2011) 56 (arXiv: 1005.1232 [gr-qc]).
\bibitem{intro19} J. E. Madriz-Aguilar and M. Bellini, JCAP {\bf 1011} (2010) 020.
\bibitem{GW1} M. Maggiore, Phys. Rep. {\bf 331} (2000) 283.
\bibitem{pc} T. Shiromizu, D. Ida and T. Torii, JHEP {\bf 11} (2001) 010.
\bibitem{Bunch} T.S. Bunch and P.C.W. Davies, Proc. R. Soc. London Ser. {\bf A 360} (1978) 117.
\bibitem{AAraya} A. Raya, J. E. Madriz-Aguilar, M. Bellini, Phys. Lett. {\bf B 638} (2006) 314.
\bibitem{BCNO} K. Bamba, S. Capozziello, S. Nojiri, S. Odintsov.
{\em Dark energby cosmology: the equivalent description via
different theoretical models and cosmography thests}. E-print:
arXiv 1205.3421.
\bibitem{AB} M. Anabitarte, M. Bellini. Eur. Phys. J. Plus {\bf
126} (2011) 90.
\bibitem{SSZ} L. Senatore, E. Silverstein, M. Zaldarriaga. {\em
New sources of gravitational waves during inflation}. E-print:
arXiv: 1109.0542.
\bibitem{D} J. Dunkley {\em et al}. Astrophys. J. Suppl. {\bf 180}
(2009) 306.
\bibitem{ur1} J. E. Madriz-Aguilar, Phys. Lett. {\bf B 645} (2007) 6-11.
\bibitem{ur2} J. E. Madriz-Aguilar, M. Anabitarte, M. Bellini, Phys. Lett. {\bf B 632} (2006) 6.
\bibitem{BM} W. Zhao, D. Baskaran and L. P Grishchuk, Phys. Rev. {\bf D 82} (2010) 043003.
\end{thebibliography}
\end{document}